\documentstyle[aps,preprint]{revtex}

\begin{document}

\draft
\tightenlines

\title{Paths to Self-Organized Criticality}
			    
\author{Ronald Dickman$^{1}$, Miguel A. Mu\~noz$^{2}$,
Alessandro Vespignani$^{3}$, and
Stefano Zapperi$^{4}$ 
}

\address{
$^1$ Departamento de F\'{\i}sica, ICEx,
Universidade Federal de Minas Gerais,
Caixa Postal 702,
30161-970 Belo Horizonte, MG, Brazil\\
$^2$  Institute {\it Carlos I} for Theoretical and Computational Physics\\
and Departamento de Electromagnetismo y F{\'\i}sica de la Materia\\
18071 Granada, Spain.\\
$^3$ The Abdus Salam International Centre for Theoretical Physics (ICTP)
P.O. Box 586, 34100 Trieste, Italy\\
$^4$PMMH - Ecole de Physique et Chimie Industrielles,
10, rue Vauquelin, 75231 Paris CEDEX 05, France \\
}
\date{\today}

\maketitle
\begin{abstract}
We present a pedagogical introduction to self-organized 
criticality (SOC), unraveling its connections with
nonequilibrium phase transitions. 
There are several paths from a conventional 
critical point to SOC.  
They begin with an absorbing-state phase
transition (directed percolation is a familiar example), and impose
supervision or driving on the system; two commonly used 
methods are extremal dynamics, and driving at a rate approaching zero.  We illustrate 
this in sandpiles, where SOC is a consequence of slow driving in a system 
exhibiting an absorbing-state phase transition with a conserved density.   
Other paths to SOC, in driven interfaces, the Bak-Sneppen model, and self-organized 
directed percolation, are also examined. We review the status of experimental realizations 
of SOC in light of these observations.
\end{abstract}

%\begin{multicols}{2}
%\narrowtext

\date{\today}

\newpage

\section{Introduction}

The label ``self-organized" is applied indiscriminately 
in the current literature to
ordering or pattern formation amongst many interacting units.
Implicit is the notion that the phenomenon of interest, be it scale invariance,
cooperation, or supra-molecular organization (e.g., micelles),
appears spontaneously.  That, of course, is just how the magnetization appears
in the Ising model; but we don't speak of ``self-organized magnetization."
After nearly a century of study, we've come to expect the spins to organize;
the zero-field magnetization below $T_c$ is no longer a surprise.
More generally, spontaneous organization of interacting units is precisely
what we seek, to explain the emergence of order in nature.  We can expect
many more surprises in the quest to discover what kinds of order a given
set of interactions lead to.  All will be {\it self}-organized, there being
no outside agent on hand to impose order!

``Self-organized {\it criticality}" (SOC) carries
greater specificity, because criticality usually does not happen spontaneously:
various parameters have to be tuned to reach the critical point.
Scale-invariance in natural systems, far from equilibrium, isn't explained
merely by showing that the interacting units {\it can} 
exhibit scale invariance at
a point in parameter space; one has to show how the system is {\it maintained}
(or maintains {\it itself}) at the critical point.  
(Alternatively one can try to show that there is
{\it generic} scale invariance, that is, 
that criticality appears over a region of parameter 
space with nonzero measure \cite{ggrin,grin91}.)  
``SOC" has been used to describe spontaneous 
scale invariance in general; 
this would seem to embrace random walks, as well as
fractal growth \cite{frac}, 
diffusive annihilation ($A$ $+$ $A$ $\rightarrow 0$
and related processes), and nonequilibrium surface 
dynamics \cite{barabasi}.  Here we restrict
the term to systems that are attracted to a critical
(scale-invariant) {\it stationary} state; 
the chief examples are sandpile models \cite{btw}.
Another class of realizations, exemplified by the Bak-Sneppen 
model \cite{baksnep}, involve extremal dynamics
(the unit with the extreme value of a certain variable is the next to change).
We will see that in many examples of SOC,
there is a choice between global supervision (an odd state of affairs
for a ``self-organized" system), or a strictly local 
dynamics in which the rate of one
or more processes must be {\it tuned} to zero.

The sandpile models introduced by Bak, Tang and Wiesenfeld (BTW) \cite{btw},
Manna \cite{manna}, and others have attracted great interest, as the first and
clearest examples of self-organized criticality.
In these models, grains of ``sand" are injected 
into the system and are lost at the boundaries, allowing
the system to reach a stationary state with a 
balance between input and output. 
The input and loss processes are linked in a special way 
to the local dynamics,
which consists of activated, conservative, 
redistribution of sand.
In the limit of infinitely slow input,
the system displays a highly fluctuating, scale-invariant avalanche-like pattern of activity.
One may associate rates $h$ and $\epsilon$, respectively, with the 
addition and removal processes.  
We have to adjust these parameters to realize SOC: 
it appears in the limit of 
$h$ and $\epsilon \rightarrow 0^+$ with
$h/\epsilon \rightarrow 0$ \cite{ggrin,hwa,vz,dvz}.  
(The addition and removal processes occur infinitely
slowly compared to the local redistribution dynamics, 
which proceeds at a rate of unity.
Loss is typically restricted to the boundaries, 
so that $\epsilon \rightarrow 0$
is implicit in the infinite-size limit.)

Questions about SOC fall into two categories.  
First, Why does self-organized criticality
exist?  What are the conditions for a model to have SOC?  Second, the many
questions about critical 
behavior (exponents, scaling functions, power-spectra, etc.) 
of specific models, and whether these 
can be grouped into universality classes, as for conventional 
phase transitions both in and out of equilibrium.  
Answers to the second type of question come from
exact solutions \cite{dharex}, simulations \cite{sim}, 
renormalization group analyses \cite{rg},
and (one may hope) field theoretical analysis.
Despite these insights, assertions in the literature about
spontaneous or parameter-free criticality have tended to 
obscure the nature of the phase
transition in sandpiles, fostering the impression that SOC is a
phenomenon {\it sui generis}, inhabiting a different world than that of 
standard critical phenomena.
In this paper we show that SOC is a phase transition to an absorbing state,
a kind of criticality that has been well studied, principally in the guise of
directed percolation \cite{kinzel}.  
Connections between SOC and an underlying conventional phase
transition have also been pointed out by Narayan and 
Middleton \cite{nm}, and by
Sornette, Johansen and Dornic \cite{sornette}.

Starting with a simple example (Sec. II), we will see that the 
absorbing-state transition provides the mechanism for SOC (Sec. III).  
That is, we explain
the existence of SOC in sandpiles on the 
basis of a conventional critical point.  In Sec. IV
we discuss the transformation of a conventional phase transition to SOC in the 
contexts of driven interfaces, a stochastic process that reproduces the
stationary properties of directed percolation, and the Bak-Sneppen model.  We find that
criticality requires tuning, or equivalently, an infinite time-scale separation.  With this
essential point in mind, we present a brief review of the relevance of SOC models
to experiments in Sec. V.  
Sec. VI presents a summary of our ideas.
We note that this paper is not intended as a complete review
of SOC; many interesting aspects of the field are not discussed.

\section{A simple example}

We begin with a simple model of {\it activated random walkers} (ARW).  Each site $j$ of a 
lattice (with periodic boundary conditions) harbors a number $z_j = 0, 1, 2...$ of random
walkers.  (For purposes of illustration the ring $1,...,L$ will do.)  
Initially, $N$ walkers are distributed randomly amongst the sites. 
Each walker moves independently, without bias, to one of the neighboring sites 
(i.e., from site $j$ to $j+1$ or
$j-1$, with site $L+1 \equiv 1$ and $0 \equiv L$), the only restriction being that an
isolated walker (at a site with $z_j = 1$) is paralyzed until such time as another walker
or walkers joins it.  The active sites (with $z_j \geq 2$) follow a Markovian (sequential) 
dynamics: each active site loses, at a rate 1, a pair of walkers, which jump independently to
one of the neighbors of site $j$.  (Thus in one dimension there is a probability of 1/2
that each neighbor gains one walker, while with probability 1/4 both walkers hop to the left,
or to the right.)

The model we have just defined is characterized by the number of lattice sites, $L^d$, and the
number of particles, $N$.  It has two kinds of configurations: active, in which at least
one site has two or more walkers, and absorbing, in which no site is multiply occupied,
rendering all the walkers immobile \cite{mnote}.  For $N > L^d$ only active configurations are 
possible, and
since $N$ is conserved, activity continues forever.  For $N \leq L^d$ there are both active and
absorbing configurations, the latter representing a shrinking fraction of 
configuration space as the density $\zeta \equiv N/L^d \rightarrow 1$.  Given that we start in 
an active configuration (a virtual certainty for an initially random distribution with $\zeta 
>0$ 
and $L$ large), will the system remain active indefinitely, or will it fall into an absorbing 
configuration?  For
small $\zeta$ it should be easy for the latter to occur, but it seems 
reasonable that for sufficiently large densities (still $ < 1$),
the likelihood of reaching an absorbing configuration becomes so 
small that the walkers remain active
indefinitely.  In other words, we expect sustained activity for densities greater than
some critical value $\zeta_c$, with $\zeta_c < 1$.

A simple mean-field theory provides a preliminary check of this intuition.  
Consider activated random walkers
in one dimension.  For a site to gain particles, it must have an active
($z \geq 2$) nearest neighbor.  Since active sites release a pair of walkers at a rate of unity,
a given site receives a single walker from an active neighbor at rate 1/2, 
and a pair of walkers at rate 1/4.  
Thus the rate of transitions that take $z_j$ to $z_j +1$
is $[P(z_j,z_{j+1} \geq 2) + P(z_j,z_{j-1} \geq 2)]/2$; transitions from 
$z_j$ to $z_j + 2$ occur at half this rate.  In the mean-field
approximation we ignore correlations between different sites, and factorize the joint 
probability into a product:
$P(z,z' \geq 2) = \rho_z \rho_a$, where $\rho_z$ is the fraction of sites with occupation
$z$ and $\rho_a = \sum_{z\geq 2} \rho_z$ is the fraction of active sites.  Using this 
factorization,
we can write a set of equations for the site densities:

\begin{equation}
\frac{d \rho_z}{dt} = \rho_a (\rho_{z-1} - \rho_z) + \frac{1}{2} \rho_a (\rho_{z-2} - \rho_z) 
                             + \rho_{z+2} - \theta_{z-2} \rho_z \;, \;\;\;\; (z = 0, 1, 2...),
\label{arwmft}
\end{equation}
where $\theta_n = 0$ for $n < 0$ and is one otherwise.  The final two terms represent
active sites losing a pair of walkers.  It is easy to see that the total probability, and the
density $\zeta = \sum_{z} z\rho_z$ are conserved by the mean-field equations.
This infinite set of coupled equations can be integrated numerically if we
impose a cutoff at large $z$.  (This is justified by the finding that $\rho_z$ decays
exponentially for large $z$.)  The mean-field theory predicts a {\it continuous phase 
transition} 
at $\zeta_c = 1/2$.  For $\zeta < \zeta_c$ the only stationary state is the absorbing 
one, $\rho_a = 0$, while for
$\zeta \stackrel > ~ \zeta_c$ the active-site density grows $\propto \zeta - \zeta_c$.
A two-site approximation (in which we write equations for the fraction $\rho_{z,z'}$ of 
nearest-neighbor pairs with given heights, but factorize joint probabilities 
involving three or more sites), yields $\zeta_c = 0.75$.

The existence of a continuous phase transition is confirmed in Monte Carlo simulations,
which yield $\zeta_c \simeq 0.9486$ in one dimension, and $\zeta_c \simeq 0.7169$
in two dimensions.  Figure 1 shows how the stationary density of active sites $\rho_a$ depends
on $\zeta$; we see $\rho_a$ growing continuously from zero at $\zeta_c$.
(The points represent estimated densities for $L \rightarrow \infty$, based on simulation
data for $L$ = 100 --- 5000.)  The inset shows that the active-site density follows
a power law, $\rho_a \sim (\zeta -\zeta_c)^{\beta}$, with $\beta = 0.43(1)$; a finite-size
scaling analysis confirms this result \cite{snote}.
In summary, activated random walkers exhibit a continuous phase transition from an 
absorbing to an active state as the particle density is increased above $\zeta_c$, with
$\zeta_c$ strictly less than 1.  (It has yet to be shown rigorously that the active-site
density in the ARW model is singular at $\zeta_c$, in the infinite-size limit; 
our numerical results are fully
consistent with the existence of such a singularity.)

\subsection{Absorbing-State Phase Transitions}

Absorbing-state phase transitions are well known in condensed 
matter physics, and population and
epidemic modeling \cite{reviews}.  
The simplest example, which may be thought of as the ``Ising model" 
of this class of systems, is the {\it contact process} \cite{harris}.  
Again we have a lattice of $L^d$ sites,
each of which may be occupied (active) or vacant.  Occupied sites turn vacant at a rate of 
unity;
vacant sites become occupied at a rate of $(\lambda/2d) n_o$ 
where $n_o$ is the number of occupied
nearest neighbors (the factor $2d$ represents the number of nearest neighbors).
There is a unique absorbing configuration: all sites vacant.  For
$\lambda$ sufficiently small, the system will 
eventually fall into the absorbing state, while for
large $\lambda$ an active stationary state can be maintained.  Letting $\rho$ represent the
density of occupied sites, the mean-field theory analogous to the one formulated above for 
activated random walkers reads:

\begin{equation}
\frac{d \rho}{dt} = (\lambda -1) \rho - \lambda \rho^2 \;.
\label{cpmft}
\end{equation}
This predicts a continuous phase transition (from $\rho \equiv 0$ to $\rho = 1 -\lambda^{-1}$
in the stationary state) at $\lambda_c = 1$.  Rigorous analyses \cite{liggett,bez} 
confirm the existence of
a continuous phase transition at a critical value $\lambda_c$, in any dimension $d \geq 1$.
Simulations and series analyses yield $\lambda_c = 3.29785(2)$ in one dimension.
This model, and its continuous-update counterpart, {\it directed percolation} 
(DP; see Sec. IV), have been
studied extensively.  The critical exponents are known to good precision for $d=1$, 2, and 3;
the upper critical dimension $d_c = 4$.  There is, in addition, a
well established field theory for this class of models \cite{janssen,cardy}:

\begin{equation}
\frac{ \partial \rho}{\partial t} = \nabla^2 \rho -a \rho - b\rho^2 + \eta(x,t) \;.
\label{cpft}
\end{equation}

\noindent Here $\rho(x,t)$ is a local particle density, and $\eta (x,t)$ is a Gaussian noise 
with
autocorrelation

\begin{equation}
\langle \eta(x,t) \eta(x',t') \rangle = \Gamma \rho(x,t) \delta(x-x') \delta(t-t') \;.
\label{noisecp}
\end{equation}
That $\langle \eta^2 \rangle $ is {\it linear} in the local density follows from the
fact that the numbers of events (creation and annihilation) in a given region are Poissonian
random variables, so that the variance equals the expected value.  
(The noise must vanish when $\rho = 0$ for the latter to be an absorbing state!)
This field theory serves
as the basis for a strong claim of universality \cite{janssen,gr82}: 
{\it Continuous phase transitions to an
absorbing state fall generically in the universality class of directed percolation.}
(It is understood that the models for which we expect DP-like behavior have short-range
interactions, and are not subject to special symmetries or conservation laws beyond
the simple translation-invariance of the contact process.  Models subject to a conservation law are known to have a different critical behavior \cite{baw}.)

The activated random walkers model resembles the contact process 
in having an absorbing-state phase transition.
We should note, however, two important differences between the models.
First, ARW presents an infinite number ($2^{L^d}$, to be more precise) of absorbing 
configurations, while the CP
has but one.  In fact, particle models in which the number of
absorbing configurations grows exponentially with the system size have also been studied
intensively.  The simplest example is the {\it pair contact process}, in which both elementary 
processes (creation
and annihilation) require the presence of a nearest-neighbor pair of particles \cite{pcp1}.
In one dimension, a pair at sites $i$ and $i+1$ can either annihilate, at rate $p$, or
produce a new particle at either $i-1$ or $i+2$, at rate $1-p$ (provided the selected site is 
vacant).  This model
shows a continuous phase transition from an active state for $p < p_c$
to an absorbing state above $p_c$.  The static critical behavior again belongs to the DP 
universality class, but the
critical exponents associated with spreading of activity 
from an initially localized region are nonuniversal, varying
continuously (in one dimension) with the particle density
in the surrounding region \cite{pcp2}.

A second important difference between ARW and the CP and PCP 
is that the former is subject to a conservation law
(the number of walkers cannot change from its initial value).  
In a field-theoretic description of ARW we will
therefore need (at least) two fields: the local density $\rho (x,t)$ 
of active sites, and the local particle density
$\zeta (x,t)$; the latter is frozen in regions where $\rho = 0$.  
The evolution of $\rho$ is coupled to $\zeta$ because 
the particle density controls existence and level of activity in the ARW model.

Given that absorbing-state phase transitions fall generically in the universality class
of directed percolation, it is natural to ask whether this is the case for activated
random walkers as well.  The answer, apparently, is ``No."
The critical exponent $\beta $ for ARW is, as we noted above, 0.43, while for 
one-dimensional DP $\beta = 0.2765$ \cite{iwandp}; the other critical exponents
differ as well \cite{snote}.  While the reason for this difference is not understood,
it appears, at least, to be consistent with the existence of a conserved field
in ARW.

To summarize, our simple model of activated random walkers has 
an absorbing-state phase transition, as does the
contact process, directed percolation and the PCP.  All possess the
same basic phase diagram: active and inactive phases separated 
by a continuous phase transition at a critical value
of a ``temperature-like" parameter ($\zeta$ in ARW, $\lambda$
in the CP).  But ARW possesses an
infinite number of absorbing configurations, and the evolution 
of its order parameter (the active-site density) is
coupled to a conserved density $\zeta$.   The latter presumably
underlies its belonging to a different universality class than DP.

\section{Activated Random Walkers and Sandpiles}

The activated random walkers model possesses a conventional critical point:
we have to tune the parameter $\zeta$ to its critical value.  What 
has it got to do with {\it self-organized}
criticality?  The answer is that ARW has essentially the same 
local dynamics as a model known to exhibit SOC,
namely, the Manna sandpile \cite{manna}.  In Manna's sandpile, the redistribution dynamics
runs in parallel: at each time step, all of the sites with $z \geq 2$ simultaneously liberate
two walkers, which jump randomly to nearest neighbor sites.  This may result in a new set of
active sites, which relax at the next time step, and so on.  (Time advances by 
one unit at each lattice update, equivalent
to the unit relaxation rate of an active site in ARW.)
We defined ARW with sequential dynamics as this makes it a Markov process 
with local transitions in configuration space, like a kinetic Ising model.  There is of course 
nothing wrong in defining ARW with parallel dynamics; it too has an absorbing-state phase 
transition.

There is a much more fundamental difference between the
Manna sandpile and the ARW model: the former allows addition and loss of walkers.  
Recall that
we defined the ARW with periodic boundary conditions; walkers can never 
leave the system.  In the sandpile walkers may exit from one of the
boundary sites.  (On the square lattice, for example, a walker at an edge site has a 
probability of 1/4 to leave the system at the next step.)  If we allow 
walkers to leave, then eventually the system will
reach an absorbing configuration.  When this happens, we add a new walker 
at a randomly chosen site.   This
innocent-sounding prescription --- add a walker when and only when all other activity ceases --- 
carries the infinite time scale separation essential to the appearance
of SOC in sandpiles.  The sequence of active configurations between two successive additions
is known as an {\it avalanche}; avalanches may involve any number of sites, from zero 
(no topplings) up to the entire system.

Manna showed that his model reaches a stationary state in which
avalanches occur on all scales,
up to the size of the system, and follow a power-law distribution, $P(s) \sim s^{-\tau}$, for
$s \ll s_c$.  (Here $s$ is the number of transfer
or toppling events in a given avalanche, and $s_c \sim L^D$ is a cutoff associated with the
finite system size.)  In other words, the Manna sandpile, like the models devised by Bak, Tang
and Wiesenfeld and others, exhibits scale invariance in the stationary state.

We know that ARW, which has the same 
local dynamics as the Manna
sandpile, shows scale invariance when (and only when) the density $\zeta = \zeta_c$.
So in the stationary state of the Manna model, the density is somehow attracted to its
critical value.  How does it happen?  The mechanism of SOC depends upon a particular 
relation between the input and loss processes, and the conventional absorbing-state 
phase transition in the model with a fixed number of particles.  Walkers cannot
enter the system while it is active, though they may of course leave upon reaching 
the boundary.  In the presence of activity, then, $\zeta > \zeta_c$ and $d \zeta/dt < 0$.  
In the absence of activity there is addition, but no loss of walkers,
so $\zeta < \zeta_c$ implies $d \zeta /dt >0$.  Evidently, the only possible stationary 
value for the density in the sandpile is $\zeta_c$!
Of course, it is possible to have a low level of activity locally, in a region with
$\zeta < \zeta_c$, but under such conditions activity cannot propagate or be sustained.
(One can similarly construct absorbing configurations with $\zeta > \zeta_c$, but these are
unstable to addition of walkers, or the propagation of activity from outside.)
In the infinite-size limit, the stationary activity density is zero for $\zeta <\zeta_c$,
and positive for $\zeta > \zeta_c$, ensuring that $\zeta$ is pinned at $\zeta_c$,
when loss is contingent upon activity, and addition upon its absence.

That the Manna sandpile, in two or three dimensions, with parallel dynamics, has
a scale-invariant avalanche distribution is well known \cite{manna}.  Here we note that
the same holds for the one-dimensional version, with random sequential dynamics.
Figure 2 shows the probability distribution for the avalanche size (the total number 
of topplings) when we modify ARW
to include loss of walkers at the boundaries, and addition at a randomly chosen site, when
the system falls into an absorbing configuration.  The distribution follows a power law,
$P(s) \sim s^{-\tau_s}$, 
over a wide range of avalanche sizes and durations; there is, as expected, an exponential
cutoff $s_c \sim L^D_s$
for events larger than a characteristic value associated with the finite size of the
lattice. (Our best estimates are $\tau_s = 1.10(2)$ and $D$ = 2.21(1).)  
The upper inset of Fig. 2 shows that the stationary density
approaches $\zeta_c$, the location of the absorbing-state phase transition, as 
$L \rightarrow \infty$.  It is also interesting to note that, in contrast with certain
deterministic one-dimensional sandpile models \cite{ali,kadanoff}, the present example
appears to exhibit finite-size scaling, as shown in the lower inset of Fig. 2.

\subsection{A Recipe for SOC}

The connection between activated random walkers and the Manna sandpile 
suggests the following recipe for SOC.  Start with a system having a 
continuous absorbing-state phase transition at a
critical value of a density $\zeta$.  This density should represent the global 
value of a local dynamical
variable {\it conserved by the dynamics}.  Add to the conservative local 
dynamics (1) a process for increasing the density in infinitesimal steps
($\zeta \rightarrow \zeta + d\zeta$) when the local dynamics 
reaches an absorbing configuration, and
(2) a process for decreasing the density at an infinitesimal rate while the
system is active.  Run the system until it reaches the stationary state; 
it is now ready to display scale invariance.

Let's see how these elements operate in the Manna sandpile.  We 
started with activated random walkers,
which does indeed display a continuous absorbing-state transition as a function the
density $\zeta$ of walkers; this density, moreover, is conserved.  To 
this we added the input of one walker ($\zeta \rightarrow \zeta + 1/L^d$
in $d$ dimensions),
when the system is inactive.  We then broke the translational symmetry 
of the ARW model to define boundary sites, 
and allowed walkers at the boundary to leave the system.  The latter implies a loss 
rate $d \zeta /dt \propto - L^{-1} \rho_b$, where $\rho_b$ is the activity density 
at the boundary sites.  The conditions of our recipe are satisfied when 
$L \rightarrow \infty$, which we needed anyway, to have a 
proper phase transition in the original model.

Now we can examine the ingredients one by one.  First, the phase transition 
in the original model should be to an {\it absorbing} state, because our input 
and loss steps are conditioned on the absence or presence of activity.  Second, 
the temperature-like parameter controlling the transition should be a conserved 
density.  So the contact process and PCP aren't suitable starting points for 
SOC, because the control parameter $\lambda$ isn't a dynamical variable.  (To
self-organize criticality in the CP, we'd have to change $\lambda$ itself,
depending on the absence of presence of activity.  But this is 
tuning the parameter by hand!)  Third, we need to change the density $\zeta$ in 
infinitesimal steps, else we will always be jumping between values above or 
below $\zeta_c$ without actually hitting the critical density.  The same thing 
will happen, incidentally, if we start out with a model that has a
{\it discontinuous} transition (with attendant hysteresis) between an active 
and an absorbing state; this yields self-organized stick-slip behavior.

The basic ingredients of our recipe are an absorbing-state phase transition, and a
method for forcing the model to its critical point, by adding (removing) particles
when the system is frozen (active).  Following the recipe, the transformation of
a conventional critical point to a self-organized one 
does not seem surprising \cite{mtak}.

\subsection{Firing the Baby-Sitter}

The reader may have noted a subtle inconsistency in the above discussion.  We rejected the 
contact process as a suitable candidate for SOC because changing the parameter 
$\lambda$ on the basis of the current state (active or frozen) amounts to tuning.  
Cannot the same be said for adding walkers in the Manna sandpile?  Somehow, a 
dynamics of walkers entering and leaving the system seems more ``natural" than 
wholesale fiddling with a parameter.  But who is going to watch for activity, to know 
when to add a particle?  A system managed by a supervisor can hardly be called 
``self-organized!"
If we want to avoid building a supervisor or baby-sitter into the model, we had better 
say that addition goes on {\it continuously}, at rate $h$, and that 
{\it SOC is realized in the limit} $h \rightarrow 0^+$ \cite{vz,dvz}.  
(The original sandpile definitions {\it have} a baby-sitter.  
Simulations, in particular, have a live-in baby-sitter to decide the next move.  
Addition at rate $h \rightarrow 0^+$ is a supervisor-free interpretation of the 
dynamics \cite{broeker}.)
In the recipe for SOC without baby-sitters, we replace addition (1) above with (1'): 
allow addition at rate $h$, {\it independent} of the state of the system, and
take $h \rightarrow 0^+$.
(There is no problem with the removal step: dissipation is associated with activity, 
which is local.)
We pay a price when we fire the baby-sitter: there is now a parameter $h$ in the model,
{\it which has to be tuned to zero}.  Evidently, sandpiles don't exhibit
{\it generic} scale invariance, but rather, scale invariance at a {\it point} in
parameter space.  This is consistent with Grinstein's definition of
SOC, which requires an infinite separation of time scales from the outset \cite{ggrin}.

\subsection{Variations}

In certain respects, our recipe allows greater freedom than was explored in the 
initial sandpile models.  There is no special reason, for example, why loss of walkers
has to occur at the boundaries.  We simply require that 
activity be attended by dissipation at an infinitesimal rate.
SOC has, indeed, been demonstrated in translation-invariant 
models with a uniform 
dissipation rate $\epsilon \rho$ when 
$\epsilon \rightarrow 0^+$\cite{vz,kiss}.  
In the original sandpile models, {\it addition} takes 
place with equal probability at any site, 
but restricting addition to a subset of the lattice will still yield SOC.

Our recipe allows a tremendous amount of freedom for the starting model; the 
only restriction is that it possess an absorbing-state critical point as 
a function of a conserved density.  
The dynamical variables can be continuous or discrete. 
The hopping process does not have to be symmetric, as in ARW.  (In fact,
{\it directed} hopping yields an exactly-soluble sandpile \cite{dirsand}.)
The model need not be defined on a regular lattice; any structure with a well 
defined infinite-size limit should do.
The dynamics, moreover, can be deterministic.  Consider a variant of the ARW 
model (on a $d$-dimensional cubic lattice) in which a site is active if it has 
$z \geq 2d$ walkers.  At each lattice update (performed here with parallel dynamics),
every active site `topples, transferring a single 
walker to each of the $2d$ nearest-neighbor sites.
In this case the only randomness resides in the initial configuration.  But the model 
again exhibits a continuous absorbing-state phase transition as we tune the number 
of walkers per site, $\zeta$.
Starting with this deterministic model, our recipe yields the 
celebrated Bak-Tang-Wiesenfeld sandpile.

As a further variation, we can even relax the condition that 
the order parameter is coupled to a conserved field \cite{socolar}. 
The price is the introduction of an 
additional driving rate. This situation is exemplified
by the forest-fire model \cite{ds92,cds94}. The model is defined 
on a lattice in which each site can be in one of three states:
empty, or occupied by a tree, either live or burning. 
Burning trees turn into empty sites, 
and set fire to the trees at nearest-neighbor sites, 
at a rate of unity. 
It is easy to recognize that burning trees are 
the active sites: any
configuration without them is absorbing.
In an infinite system, there will be a critical tree density 
that separates a phase in which fires spread indefinitely 
from an absorbing phase with no burning trees.
In a finite system we can study this critical point by fixing
the density of trees at its critical value \cite{cds96}.

So far we have no process for growing new trees.
The forest-fire propagates like an epidemic with immunity: a site
can only be active once, and there is no proper steady state \cite{dynperc}.  
As in sandpiles, to obtain a SOC state we must introduce 
an external driving field $f$ that introduces a small probability
for each tree to catch fire spontaneously. This driving field 
allows the system to jump between absorbing configurations through the 
spreading of fires. The latter, however, are completely dissipative, i.e.,
the number of trees is not conserved. Thus, if we want to reach a 
stationary state we must introduce a {\it second} external driving field $p$
that causes new trees to appear.  (Empty sites become occupied by a
living tree at rate $p$.)
In this case criticality is reached by the double slow driving condition
$f,p\to0$ and $f/p\to0$. In practice, this slow driving condition
is achieved by the usual supervisor, that stops fire ignition and tree
growth during active intervals.

\subsection{Fixed-Energy Sandpiles}

If someone hands us a sandpile displaying SOC, we can identify the initial model 
in our recipe; it has the same local dynamics as the SOC sandpile.  Thinking of the 
conserved $\zeta$ as an energy density, we call the starting model a 
{\it fixed-energy sandpile} (FES).  Thus the activated random walkers model introduced 
in Sec. II is the fixed-energy Manna sandpile, and the variant described in the preceding 
subsection is the BTW FES.  Now the essential feature of the fixed-energy sandpile is 
an absorbing-state phase transition.  SOC appears when we rig up the addition and 
removal processes to drive the local FES dynamics to $\zeta_c$.  To understand the 
details of SOC, then, we ought to try to understand the conventional phase 
transition in the corresponding fixed-energy sandpile.  
This is our program for 
addressing the second class of questions 
(about critical exponents and universality 
classes) mentioned in the Introduction.  
Since fixed-energy sandpiles have a simple 
dynamics (Markovian or deterministic) without loss or addition, and are 
translation-invariant (when defined on a regular lattice), 
they should be 
easier to study than their SOC counterparts.
The relation to absorbing-state phase transitions leads to a proper identification
of the order parameter \cite{vz}, and suggests a strategy for constructing
a field theory of sandpiles \cite{vdmz}.
Spreading exponents, conventionally measured in absorbing-state 
phase transitions, are related through scaling laws to avalanche 
exponents, usually measured in slowly driven systems \cite{br,cmv}.

\section{Other Paths to SOC}

\subsection{Driven Interfaces}

In this section we illustrate the central idea of the preceding section --- the transformation 
of a conventional phase transition to a self-organized one --- in a different, though related,
context.  We begin with a single point mass undergoing driven, dissipative motion in
one dimension.  Its position $H(t)$ follows the equation of motion 

\begin{equation}
M\frac{d^2H}{dt^2}+\gamma \frac{dH}{dt}=F-F_p(H),
\label{eq:1}
\end{equation}
where $M$ is the mass, $\gamma \dot{H}$ 
represents viscous dissipation, $F$ is the applied force, and $F_p(H)$
is a position-dependent pinning force.  In many cases 
of interest (i.e., domain walls or flux-lines)
the motion is overdamped and we 
may safely set $M=0$.  The pinning force has mean zero 
($\langle F_p (h) \rangle = 0$) and its autocorrelation
$\langle F_p(h) F_p(h+y) \rangle \equiv \Delta(|y|)$ decays 
rapidly with $|y|$; the statistical properties of $F_p$ are independent of $H$.
Assuming, as is reasonable, that $F_p$ is bounded ($F_p \leq F_M$),
we expect the motion to continue if the driving force $F$ exceeds $F_M$.  Otherwise
the particle gets stuck somewhere.

Now consider an elastic interface (or a flux line) subject
to an external force, viscous damping, and a pinning force associated with
irregularities in the surrounding medium.  If we discretize our interface, using
$H_i (t)$ to represent the position, along the direction of
the driving force, of the $i$-th segment , the equation of motion is

\begin{equation}
\gamma \frac{dH_i}{dt} = H_{i+1} + H_{i-1} - 2H_i(t) + F-F_{p,i} (H_i),
\label{eq:2}
\end{equation}
where the $F_{p,i}(H_i)$ are a set of independent pinning forces with statistical
properties as above.  This driven interface model has a {\it depinning transition}
at a critical value, $F_c$, of the driving force \cite{lesch}.  (Eq. (\ref{eq:2}) describes
a {\it linear} driven interface, so-called because it lacks the nonlinear term 
$\propto (\nabla h)^2$, familiar from the KPZ equation \cite{barabasi,kpz}.)
For $F < F_c$ the motion is eventually
arrested ($dH_i/dt = 0$ for all $i$), while for $F > F_c$ movement continues indefinitely.
Close to $F_c$ there are avalanche-like bursts of movement on all scales, interspersed
with intervals of near-standstill.  The correlation length and relaxation time diverge at
$F_c$, as in the other examples of absorbing-state phase transitions we've discussed
above.  We may take the order parameter for this transition as the mean velocity,
$\overline{v} = \langle dH_i/dt \rangle$.

To reach the absorbing-state phase transition in the driven interface model 
we need to adjust the applied force $F$ to its critical value $F_c$.  Can we modify this 
system so that it will be {\it attracted} to the critical state?  Note that $F$ is not a 
dynamical 
variable, any more
than is $\lambda$, in the contact process.  Our sandpile recipe doesn't seem
to apply here.  The crucial observation is that we may change the nature of the driving,
replacing the constant force $F$ with a constraint of
{\it fixed velocity}, $dH_i/dt = v$.  A finite $v$ corresponds to a state in the active phase:
the mean driving force
$\langle F_i \rangle_v > F_c$ for $v > 0$.  When we allow $v$ to tend to zero from above,
we approach the depinning transition.
This limit can be attained through an extremal dynamics in which we 
advance, at a 
given step, only the element subject to the 
smallest pinning force \cite{invp,sod}. 
(Notice that
in extremal dynamics we are directly adjusting 
the {\it order parameter}\cite{sornette}.)

To avoid the global supervision implicit in
extremal dynamics we may attach each element of the
interface to a spring, and move the other end of each spring at speed $V$.  Now the
equations of motion read

\begin{equation}
\gamma \frac{dH_i}{dt} = H_{i+1} + H_{i-1} - 2H_i(t) + k(Vt - H_i) - F_{p,i}(H),
\label{eq:3}
\end{equation}
where $k$ is the spring constant.
For high applied velocities,
the interface will in general move smoothly, with velocity
$\dot{H} = V$, while for low $V$ stick-slip motion is likely.
In the overdamped regime, the amplitudes of the slips are
controlled by $V$ and $k$, and the statistics of the potential.
In the limit $V \to 0$, the interface motion exhibits 
scale invariance; $V$ plays a role analogous to $h$
in the sandpile.
(The limits $V\to 0$ and $k\to 0$ have a particular
significance, since the block can explore
the pinning-force landscape quasistatically.)
The fine tuning of $F$ to $F_c$ in the constant-force driving
has been replaced by fine tuning $V$ to zero.  This parameter tuning
corresponds, once again, to an infinite time-scale separation.
Finally, we note that restoring inertia ($M >0$)
results in a discontinuous depinning transition with hysteresis,
resulting in stick-slip motion of the sort associated with friction \cite{persson}.

Once again, we have transformed an absorbing-state phase transition ($F=F_c$)
into SOC by driving the system at a rate approaching zero ($V \to 0$).
But there appear to be fundamental differences between sandpiles
and driven interfaces.  In the sandpile, but not in the driven interface, 
the order parameter is coupled to a conserved density.  The sandpile,
moreover, does not involve a quenched random field as does the
driven interface.  Despite these apparent differences, close
connections have been suggested between the two kinds of model \cite{nm,nf,pacz,lau}.
We review this correspondence in the next subsection, 
following Ref.~\cite{lau}.

\subsection{Sandpiles and Driven Interfaces}

Consider the BTW fixed-energy sandpile
in two dimensions; let $H_i (t)$ be the number of times site $i$ has toppled since time
zero. To write a dynamics for $H_i$, we observe that the occupation $z_i(t)$ of 
site $i$ differs from its initial value, $z_i(0)$,
due to the inflow and the outflow of particles at this site.
The outflow is given by $4H_i(t)$, since each toppling expels four 
particles. The inflow 
can be expressed as $\sum_{NN} H_j(t)$: site $i$ gains a particle each time
one of its nearest neighbors topples.
Summing the above contributions we obtain:
\begin{eqnarray}
z_i(t) & = & z_i(0)+\sum_{j NN i} H_j(t) - 4H_i(t) \nonumber \\
& = & z_i(0) + \nabla_D^2 H_i(t),
\label{hdiscr}
\end{eqnarray}
where $\nabla_D^2$ stands for the discretized Laplacian.
Since sites with $z_i(t) \geq 4$ topple at unit rate, the dynamics of $H_i$
is given by

\begin{eqnarray}
\frac{d H_i}{d t} & = &\Theta [z_i(0) + \nabla_D^2 H_i(t) - 3] \nonumber \\
	        & = & \Theta [\nabla_D^2 H_i(t) + F - F_{p,i}] ,
\label{hdyn}
\end{eqnarray}
where $d H_i/d t$ is shorthand for the {\it rate} at
which the integer-valued variable $H_i(t)$ jumps to $H_i(t) +1$, and
$\Theta (x) = 1$ for $x > 0$ and is zero otherwise.
In the second line,
$F \equiv \zeta - 3$ and $F_{P,i} \equiv z_i (0) - \zeta$.  
(Recall that $\zeta = \langle z_i (t) \rangle$ for all $t$.)
Thinking of $H_i (t)$ as
a discretized interface height, Eq. (\ref{hdyn}) represents an overdamped, driven
interface in the presence of {\it columnar} noise, $F_{p,i}$, which takes independent
values at each site, but does not depend upon $H_i$, as it does in the interface model
discussed in the preceding subsection.  We see from this equation that tuning
$\zeta $ to its critical value $\zeta_c$ is analogous to
tuning the driving force to $F_c$.  If we replace the discrete height $H_i$ in
Eq. (\ref{hdyn}) with a continuous field, $H(x,t)$ (and similarly for $F_p$),
and replace the $\Theta$-function by its argument,
we obtain the Edwards-Wilkinson surface-growth model with columnar disorder,
which has been studied extensively \cite{col}.
The similarity between the present height representation and the
dynamics of a driven interface suggests that the critical point
of the BTW fixed-energy sandpile belongs to the universality class of
linear interface depinning
with columnar noise, if the rather violent nonlinearity of the $\Theta$-function
is irrelevant.  (The latter remains an open question.  A height representation for the
Manna sandpile is also possible, but is complicated by the stochastic nature of
the dynamics.)

Applying the recipe of Sec. III to the driven interface,
we would impose open boundaries, which drag behind
the interior as they have fewer neighbors pulling on them; eventually the
interface gets stuck.  When this happens, we ratchet up the ``force" at a randomly
chosen site (in effect, $F_{p,j} \to F_{p,j} -1$ at the chosen site).  The dynamics is then
attracted to the critical point.  Once again, we may trade supervision (checking
if the interface is stuck) for a constant drive ($F \to F + ht$) in the limit $h \to 0$.

\subsection{Self-Organized Directed Percolation and the Bak-Sneppen Model}

Take the square lattice and rotate it by $45^o$, so that each site has two nearest neighbors
in the row above, and two below.  The sites exist in one of two states, ``wet" and ``dry."
The states of the sites in the zeroth (top) row can be assigned at will; this defines the
initial condition.  A site in row $i \geq 1$ is obliged to be dry if both its neighbors in row
$i-1$ are dry; otherwise, it is wet with probability $p$, and dry with probability $1-p$.
This stochastic cellular automaton is called {\it site directed percolation}.  
Like the contact process, it possesses an 
absorbing state: all sites dry in row $k$ implies all dry in all subsequent rows.  
The dynamics of site DP can be expressed in a compact form if we define the site variable
$x_j^i$ to be zero (one) if site $j$ in row $i$ is wet (dry).  The variables in the next row 
are given by

\begin{equation}
x_j^{i+1} = \Theta [\max \{ \eta_j^i , \min \{ x_{j-1}^i , x_{j+1}^i \} \} - p] \;,
\label{sdp}
\end{equation}
where the $\eta_j^i$ are independent random variables, uniform on [0,1].
If both neighbors in the preceding row are in state 1, $x_j^{i+1}$ must also equal 1;
otherwise $x_j^{i+1} = 0$ with probability $p$.
Thinking of the
rows as time slices, we see that site DP is a parallel-update version of the contact process:
increasing $p$ renders the survival and propagation of the wet state more probable,
and is analogous to increasing $\lambda $ in the CP.
Just as the CP has a phase transition at $\lambda_c$, site DP has a transition
from the absorbing to the active phase at $p_c \simeq 0.7054$.

We've already dismissed the contact process (and by extension DP) as starting models
for realizing SOC via the recipe of Sec. III.  Remarkably, however, it is possible to define
a parameter-free stochastic process whose stationary state reproduces the 
properties of {\it critical} DP \cite{hansen,grzhang,maszhang}.  This process, 
self-organized directed percolation (SODP),  is
obtained by replacing the discrete variables in Eq. (\ref{sdp}) by real variables which
store the value of one of the previous $\eta_j^i$.  In place of Eq. (\ref{sdp}) we have simply

\begin{equation}
x_j^{i+1} = \max \{ \eta_j^i , \min \{ x_{j-1}^i , x_{j+1}^i \} \} \;,
\label{sodp}
\end{equation}
Notice that parameter $p$ has disappeared, along with the $\Theta$ function.
Starting from a distribution with $x_j^0 < 1$ for at least one site (but otherwise arbitrary),
this process eventually reaches a stationary state, characterized by the probability
density $\mu (x)$.  One finds that $\mu (x)$ is zero for $x < p_c$ 
(the critical value of site DP),
jumps to a nonzero value (infinity, in the thermodynamic limit), at $p_c$, and decreases 
smoothly with $x$ for $x > p_c$.
The process has discovered the critical value of
site directed percolation!

Hansen and Roux explained how this works \cite{hansen}: for any $ p \in [0,1]$
the probability that $x_j^i < p$ is $p$ if either or both of the neighbors in the previous
time slice have values less that $p$ (i.e., if the smaller of 
$x_{j-1}^{i-1} $ and $x_{j+1}^{i-1}$ is $< p$),
and is zero if $x_{j-1}^{i-1} $ and $x_{j+1}^{i-1}$ both exceed $p$.  
This is exactly how the ``wet" state propagates in site DP, with parameter $p$, if we equate
the events `site $j$ in row $i$ is wet' and `$x_j^i < p$.'   
It follows that in the stationary state, 

\begin{equation}
\Pr [ x_j^i < p] = \int _0 ^p \mu (x) dx,
\end{equation}
equals the probability $P(p)$ that a randomly chosen site is wet, in the stationary state of 
site DP
with parameter $p$.  This explains why $\mu (x) = 0 $ for $x < p_c$, and why $\mu (p_c)$
is infinite in the infinite-size limit ($dP/dp$ is infinite at $p_c$).  The spatio-temporal
distribution of DP is also reproduced; for example, the joint probability
$\Pr [x_j^i \leq p_c, x_k^i \leq p_c]$ decays as a power law for large separations $|j-k|$.  The 
process effectively studies all values of $p$ at once, greatly
improving efficiency in simulations.  Stochastic processes
corresponding to other models (DP on other lattices, bond instead of
site DP, epidemic processes) have also been devised \cite{grzhang,bagnoli}.
It seems unlikely, on the other hand, that such a real-valued stochastic process exists
for activated random walkers or other fixed-energy sandpiles.  
(Of course, such a process would be of great help in studying sandpiles!)

SODP doesn't fit into the same scheme as sandpiles or
driven interfaces.  It is a real-valued stochastic process that generates, by construction,
the probability distribution of DP for {\it all} parameter values, including $p_c$.
The process itself does not have a phase transition; all sites are active (except those inside a
sequence of 1's --- a configuration that will never arise spontaneously),
since there is a finite probability for $x_j^i$ to change.  SODP is self-organized in the
sense that its stationary probability density has a critical singularity, without
the need to adjust parameters.  If we choose to regard SODP as an instance of SOC, we must
recognize that the path in this case is very different from that in sandpiles or
driven interfaces; the system is not being forced to its critical point by external
supervision or driving. Rather, SODP is directed percolation implemented in a 
different (parameter-free) way.  Furthermore,
the dynamics embodied in Eq. (\ref{sodp}) seems a much less
realistic description of a physical system than is driven-interface motion, or even
the rather artificial dynamics of a sandpile model.  In the rather unlikely event that
SODP were realized in a natural system, it would not immediately yield
a scale-invariant ``signal" such as avalanches or fractal patterns.  The latter would
require a second process (or an observer) capable of
making fine distinctions among values of $x$ in the neighborhood of $p_c$.
So the kind of SOC represented by SODP does not appear a likely explanation
of scale invariance in nature.

A (fanciful) interpretation of Eq. (\ref{sodp}) is that $x_j^i$ represents the ``fitness"
of an individual, which mates with its neighbor to produce an offspring that inherits
the fitness of the less-fit parent.  This offspring survives if her fitness exceeds that of
an interloper, whose fitness is random.  (It is, to put it crudely,
as if an established population were constantly challenged by a flux of
outsiders.)  Seen in this light, SODP bears some resemblance to the evolutionary dynamics
represented, again in very abstract form, in the Bak-Sneppen model \cite{baksnep}.  
Here, the {\it globally} minimum fitness variable, along with its nearest neighbors,
is replaced by a [0,1] random number at each time step.  (If the $x_j^i$ are associated with
different species, then the appearance  of a new 
species at site $i$ affects the fitness of the ``neighboring" species in
the community in an unpredictable way.)
This is a kind of extremal dynamics, a scheme we've already encountered in the driven
interface model; another familiar example is invasion percolation \cite{invp}.
Interestingly, the Bak-Sneppen model shows the same qualitative
behavior as SODP: a singular stationary distribution of fitness values $x_j^i$.
The model exhibits avalanches in which replacement of a single species
provokes a large number of extinctions.

In the interface under extremal dynamics, the height
$H_i(t)$ cannot decrease.  In the Bak-Sneppen
model momentary setbacks are allowed ($x_j $ can decrease in a given step),
but individuals of low fitness will eventually be culled.  This is like an interface model
with quenched noise such that, on advancing to a new position, an element
may encounter a force that throws it backward, for a net negative displacement.
The Bak-Sneppen model is equivalent to a driven interface in which
the least-stable site and its neighbors are updated at the same moment;
we can, as before, trade extremal dynamics for a limit of infinitely
slow driving. 
 
Another way of obtaining the extremal dynamics of the Bak-Sneppen model
as the limit of a stochastic process with purely local dynamics is 
as follows \cite{bstemp}.
Take a one-dimensional lattice (with periodic boundaries, for definiteness),
and assign random numbers $x_j$, independent and uniform on [0,1], to each 
site $j= 1,...,L$.  The configuration evolves via a series of ``flips," which
reset the variables at three consecutive sites.
That is, when site $j$ flips, we replace $x_{j-1}$, $x_j$, and $x_{j+1}$ with three
independent random numbers again drawn uniformly from [0,1].  Let the {\it rate}
of flipping at site $j$ be $\Gamma e^{-\beta x_j}$, where $\Gamma^{-1}$ is a
characteristic time, irrelevant to stationary properties.  The Bak-Sneppen model
is the $\beta \to \infty $ limit of this process.

We can get some insight into the stationary behavior via a simple analysis.
Let $p(x) dx$ be the probability that $x_j \in [x, x+dx]$.  The probability density
satisfies

\begin{equation}
\frac{dp(x)}{dt} = -e^{-\beta x} p(x) - 2 \int_0^1 e^{-\beta y} p(x,y) dy 
		+ 3 \int_0^1 e^{-\beta y} p(y) dy
\label{bsp}
\end{equation}
where $p(x,y)$ is the joint density for a pair of nearest-neighbor sites.  If we
invoke a mean-field factorization, $p(x,y) = p(x)p(y)$, then

\begin{equation}
\frac{dp(x)}{dt} = -p(x) \left[ e^{-\beta x} + 2 I(\beta) \right] + 3 I(\beta),
\label{bspmft}
\end{equation}
where

\begin{equation}
I(\beta) \equiv \int_0^1 e^{-\beta y} p(y) dy \;. 
\label{ibeta}
\end{equation}
The stationary solution is 

\begin{equation}
p_{st}(x) = \frac{3}{2}  \frac {1- e^{-2\beta/3}}{1 - e^{-2\beta/3} + e^{-\beta x}(e^{\beta/3} 
-1)} \;.
\end{equation}
The solution is uniform on [0,1] for $\beta =0$, as we'd expect, but in the $\beta \to \infty$ 
limit
we have $p_{st} = (3/2) \Theta(x-1/3) \Theta(1-x)$.  The probability density 
develops a step-function singularity, as
in the Bak-Sneppen model.  Not surprisingly, the mean-field approximation yields a rather
poor prediction for the location of the singularity, which actually falls at 0.6670(1) 
\cite{grbs}.  (A two-site approximation places the singularity at $x=1/2$.)
The main point is that to realize singular behavior from a local dynamics,
we have to tune a parameter associated with the rates.  Alternative mean-field
treatments of the Bak-Sneppen model may be found in Refs. \cite{fsb} and
\cite{pmb}

We can construct a model with the 
same local dynamics as that of 
Bak and Sneppen by replacing $x_{j-1}$, $x_j$, and $x_{j+1}$ at 
rate 1, if and only if $x_j <r$.  
(Sites with $x_j >r$ may only change if they have a nearest neighbor 
below the cutoff.)  
In other words, only sites with $x_j <r$ are active; an updated site
is active with probability $r$.
There is an absorbing phase for small $r$, separated from an active phase by
a critical point at some $r_c$ \cite{pmb,jov,lipo}.   
To get the Bak-Sneppen model we
forget about $r$, and declare the unique active site in the 
system to be the one
with the smallest value of $r$.  
In the infinite-size limit, the probability to find a site
with $r < r_c$ is zero, in the stationary state.  We see once again that
in extremal dynamics we tune the order parameter itself
to zero: at each instant there is exactly one active site, so $\rho_a = 1/L$.

Grassberger and Zhang observed that the existence of SODP
``casts doubt on the significance of self-organized as opposed to ordinary
criticality."  A similar doubt might be prompted by our recipe for turning
a conventional critical point self-organized.  Of course, even if it is possible
to explain all instances of SOC
in terms of an underlying conventional critical point, the details of the
critical behavior remain to be understood \cite{tnote}.
Numerical results indicate that sandpiles, driven interfaces, and the Bak-Sneppen model
define a series of new universality classes.  Furthermore, no one has been able to
derive the critical exponents of avalanches in SOC sandpiles, even the in abelian case,
where quite a lot is known about the stationary properties \cite{anote}.

\section{SOC and the Real World}
 
Since SOC has been claimed to be the way ``nature works'' \cite{pb}, we
would expect to find a multitude of experimental examples
where this concept is useful.
Originally, SOC was considered an explanation of power laws,
that it provided a means whereby a system could
self-tune its parameters. So once we saw a power law  we could
claim that it was self-generated and ``explained'' by SOC. 
The previous sections should have convinced the
reader that there are no self-tuning critical points, although
sometimes the fine tuning is hidden, as in sandpile
models. Therefore, an ``explanation'' of experimentally
observed power laws requires the identification
of the tuning parameters controlling the scaling, as in any
other ordinary critical point. 

Here, we will restrict the discussion to experimental examples 
of avalanche behavior, leaving aside fractals and $1/f$ noise
whose connection with SOC is rather loose.  
(It is worth mentioning that a physical realization of
self-organized criticality --- without avalanches, as far as is known ---
has been identified in liquid $^4$He at the $\lambda$ point \cite{machta}.)
Following the introduction of SOC, there were many
experimental studies of avalanches, which
sometimes yielded power-law distributions over a few
decades, leading to endless discussions about the applicability
of SOC. If we accept that self-tuned critical points don't exist,
then these controversies have no basis: we have only to
understand how far the system is from the critical point, and why. 
This task has only been accomplished in a few
cases; several examples require further study,
both experimental and theoretical.

Soon after the sandpile model was introduced, 
several experimental groups measured the 
size-distribution of avalanches in granular materials.
Unfortunately, real sandpiles do not seem to 
be behave as the SOC sandpile model. Experiments  
show large periodic avalanches separated by quiescent
states with only limited activity \cite{sand}. While for small
piles one could try to fit the avalanche distribution
with a power law over a limited range \cite{sand2}, 
the behavior would eventually cross over, on increasing the system size,
to the one described above, which is not scale-invariant. 
The reason sand does not behave like an ideal sandpile
is the inertia of the rolling grains.
As grains are added, the inclination of the pile increases until it reaches
the angle of maximal stability $\theta_s$, at which point grains
start to flow.  Due to inertia, the flow does not stop
when the inclination falls to $\theta_c$, but continues until 
the inclination attains the angle of repose 
$\theta_s<\theta_c$ \cite{nagel}. Since the ``constant force''
(i.e., with $\theta$ controlled) version of the system 
has a {\it first-order} transition, it is no wonder that
criticality is not observed in the slowly driven case.
So if we want to see power-law avalanches we have
to get rid of the inertia of the grains. Grains with small
inertia exist and can be bought in any grocery store:
rice! A ricepile was carefully studied in Oslo: elongated 
grains poured at very small rate gave rise to a
convincing power-law avalanche distribution \cite{rice}. 

The previous discussion tells us that in order to observe a
power-law avalanche distribution, inertia should be negligible.
As discussed in Sec. IV, the motion of domain walls
in ferromagnets and flux lines in type II superconductors
is overdamped, due to eddy-current dissipation; these systems
are probably the cleanest experimental example of 
power-law distributed avalanches. 
The noise produced by domain wall motion is known as
the Barkhausen effect, first detected in 1919 \cite{bark}.
Since then, it has become a common non-destructive
method for testing magnetic materials, and its statistical
properties have been studied in detail. When the external
magnetic field is increased slowly, it is possible
to observe well separated avalanches, whose size 
distribution is a power-law over more than three decades 
\cite{bdm,durin1,sava,urbach,ZAP-98}.
Domain walls are pushed through
a disordered medium by the magnetic field, so we would expect a depinning transition
at some critical field $H=H_c$. One should note, however, 
that the ``internal field''
acting on the domains is not the external field, but 
is corrected by the demagnetizing field $H_d \simeq -N M$
where $M$ is the magnetization \cite{urbach,ZAP-98}
and $N$ the demagnetizing factor. 
Therefore, if we increase the external field at constant 
rate $c$, the internal  field is given by 
$H_{int} = ct-NM = ct-ky(t)$,
where $y(t)$ is the average position of the domain wall
and $k\propto N$.
We recognize here the recipe for SOC given
in section IVA: in the limit $c\to 0$ and $k\to 0$ we
expect to reach the critical point. This fact was indeed
verified in experiments, where $k$ can be controlled by
modifying the aspect ratio of the sample \cite{ZAP-98}. 

In type II superconductors, when the external field
is increased, flux lines are nucleated at the border of the sample
and pushed inside by their mutual repulsion.
The resulting flux density gradient, known as the Bean
state \cite{bean}, bears some analogy
with sandpiles, as pointed out by De Gennes over 30 years ago
\cite{degennes}. Unlike sand grains, flux lines
have little inertia, and exhibit power-law distributed avalanches \cite{flux}. 
It is still unclear whether in this
system a mechanism similar to the demagnetizing field
maintains a stationary avalanche state,
as in ferromagnets. Simulations of flux
line motion \cite{nori} have reproduced experimental results in part,
but a complete quantitative explanation of the phenomenon
is lacking.

Another broad class of phenomena where SOC has been
invoked on several occasions is that 
of mechanical instabilities: fracture, plasticity and
dislocation dynamics. Materials subject to an external
stress release acoustic signals that are often distributed
as power laws over a limited range: examples are
the fracturing of wood \cite{ciliberto}, cellular glass
\cite{strauven} and concrete \cite{ae}, in hydrogen precipitation
\cite{ccc}, and in dislocation motion in ice crystals \cite{grasso}.
While it has often been claimed that these experiments
provided a direct evidence of SOC, this is far from being established.
In fact, fracture is an irreversible phenomenon
and often the acoustic emission increases with the applied 
stress \cite{ciliberto} with a sharp peak at the failure point.
There is thus no stationary state in fracture, and
it is debated whether the failure point can even be described
as a critical point \cite{sor} or a first-order transition
\cite{zrsv}. The situation might be different in plastic deformation,
where a steady state is possible \cite{zvs}; recent
experimental measurements of dislocation 
motion appear promising \cite{grasso}.
We may mention some related phenomena in which avalanches
have been observed, and a theoretical interpretation
is still debated: martensitic transformations
\cite{mart}, sliding systems \cite{sliding} 
and sheared foams \cite{foam}.

Finally, it is worth mentioning that SOC has been claimed
to apply to several other situations in geophysics,
biology and economics. We have deliberately 
chosen to discuss only those examples for which experimental 
observations are accurate
and reproducible. Even in these cases, it is often hard
to distinguish between SOC-like behavior and other
mechanisms for generating power laws. This task appears
almost hopeless in situations where only limited data sets
are available, such as for forest fires \cite{forest},
or evolution \cite{evolution},
and remains very complicated in other cases, such as
earthquakes, as witnessed by the vast theoretical literature
on the subject \cite{earth}.

\section{Summary}

The genesis of self-organized criticality is a continuous absorbing-state phase transition.
The dynamical system exhibiting the latter may be continuous or discrete, deterministic
or stochastic, conservative or dissipative.  To transform a conventional phase transition
to SOC, we couple the local dynamics of the dynamical system to an external supervisor, or to a
``drive" (sources and sinks with rates \{$h$\}).  The relevant parameter(s) \{$\zeta$\} 
associated with the phase transition are controlled by the supervisor or drive, {\it in a way 
that does not make explicit reference} to \{$\zeta$\}.  One such path involves
slow driving ($h \to 0$), in which the interaction with the environment is contingent
on the presence or absence of activity in the system (linked to \{$\zeta$\} via the 
absorbing-state phase transition).  Another, extremal dynamics,
restricts activity to the least stable element
in the system, thereby tuning the order parameter itself to zero.  Specific realizations of this
rather abstract (and general) scheme have been discussed in the preceding sections: sandpiles,
forest fires, driven interfaces, and the Bak-Sneppen model.

Viewed in this light, ``self-organized criticality" refers neither to spontaneous or 
parameter-free criticality, nor to self-tuning. It becomes, rather, a useful concept for 
describing systems that, in
isolation, would manifest a phase transition between active and frozen regimes, and that 
are in fact driven slowly from outside.
\vspace{1em}

\noindent {\bf Acknowledgements}
\vspace{1em}

We thank M. Alava, A. Barrat, A. Chessa, D.Dhar, P.L. Garrido, P. Grassberger, D. Head,
K.B. Lauritsen, J. Machta, E. Marinari, R. Pastor-Satorras, L. Pietronero and A.Stella for
continuous discussions and fruitful ``arguments"  on the significance of SOC. 
M.A.M., A.V., and S.Z. Acknowledge partial support from the 
European Network Contract No. ERBFMRXCT980183.  M.A.M. also acknowledges
support from the Spanish Ministerio de Educaci\'on
under project DGESEIC, PB97-0842'.

\newpage
\noindent Figure Captions
\vspace{1em}

\noindent Fig. 1. Stationary density $\rho$ of active sites versus density of walkers
$\zeta$ in one-dimensional ARW.  The inset is a logarithmic plot of the same data,
where $\Delta = \zeta - \zeta_c$.  The slope of the straight line is 0.43.
\vspace{1em}

\noindent Fig. 2. Stationary avalanche-size distribution in the one-dimensional Manna
sandpile with sequential dynamics, for $L = 500$, 1000, 2000, and 5000 (left to right) . 
Lower inset: finite-size scaling plot of the data in the main graph,
$\ln P^*$ versus $\ln s^*$, with $s^* \equiv L^{-2.21} s$ and $P^* \equiv L^{2.43} P$.
Upper inset: stationary density $\zeta$ in the inner 10\% of the system, plotted versus
$1/L$.  The diamond on the $\zeta$ axis is the critical density of ARW.


\begin{references}
\bibitem{mnote}
	The expert reader will have noticed that activated random walks follow the
	same local dynamics as the Manna sandpile \cite{manna}.
\end{references}
\end{document}